\documentclass[aps,amsmath,showpacs,amsfonts,10pt]{revtex4}
\usepackage{epsfig,graphicx}
\usepackage[english]{babel}
\usepackage{amsfonts}
\usepackage{amsmath}
\usepackage{latexsym}
\usepackage{graphics,bm}
\usepackage{natbib}
\usepackage{dcolumn}
\usepackage{bm}
\usepackage{rotating}

\begin{document}

\title{Excitations of optomechanically driven Bose-Einstein condensates in a cavity: photodetection measurements}

\author{Neha Aggarwal$^{1,2}$, Sonam Mahajan$^{1}$, Aranya B. Bhattacherjee$^{2,3}$ and Man Mohan$^{1}$}

\address{$^{1}$Department of Physics and Astrophysics, University of Delhi, Delhi-110007, India} \address{$^{2}$Department of Physics, ARSD College, University of Delhi (South Campus), New Delhi-110021, India}\address{$^{3}$School of Physical Sciences, Jawaharlal Nehru University, New Delhi-110067, India}

\begin{abstract}

We present a detailed study to analyse the Dicke quantum phase transition within the thermodynamic limit for an optomechanically driven Bose-Einstein condensates in a cavity. The photodetection-based quantum optical measurements have been performed to study the dynamics and excitations of this optomechanical Dicke system. For this, we discuss the eigenvalue analysis, fluorescence spectrum and the homodyne spectrum of the system. It has been shown that the normal phase is negligibly affected by the mechanical mode of the mirror while it has a significant effect in the superradiant phase. We have observed that the eigenvalues and both the spectra exhibit distinct features that can be identified with the photonic, atomic and phononic branches. In the fluorescence spectra, we further observe an asymmetric coherent energy exchange between the three degrees of freedom of the system in the superradiant phase arising as a result of optomechanical interaction and Bloch-Siegert shift.

\end{abstract}


\maketitle

\section{Introduction}

The field of optomechanics has undergone rapid development over the last decade with application in a wide variety of systems ranging from vibrating microtoroids \citep{carmon,schliesser}, membranes \citep{thompson}, nanomechanical cantilevers \citep{hohberger,gigan,arcizet,kleckner,favero,regal}, gravitational wave detectors (LIGO project) \citep{corbitt,corbitt1} and ultracold atoms \citep{brennecke,murch,bhattacherjee,bhattacherjee1,treutlein}. The classical optomechanics is a well developed field of optical engineering. In this field, the microelectromechanical systems form a vital component in high technology ranging from iPhones to sensors. Moreover, the quantum control of mechanical motion has also been achieved for the first time in ion traps \citep{leib}. The significant goal of the field of optomechanics is to cool the optomechanical systems to their ground state \citep{mancini,vitali,genes,sonam,sonam1}. Braginsky \citep{braginsky} and Caves \citep{caves} performed pioneering work in this field, in which it was anticipated that the back-action arises due to the radiation pressure force applied by the light field on the movable mirror. This radiation pressure leads to the interaction between intensity of the optical field and the displacement of the moving mirror. This interaction of optical and mechanical degrees of freedom through radiation pressure has been used in the field of gravitational wave detectors \citep{caves1,loudon} and laser cooling \citep{hansch,wineland,chu}. In the recent years, an optomechanical system containing a BEC has also been examined \citep{paternostro}.

The Dicke model has also grabbed considerable attention for decades which describes the two-level systems or spins uniformly interacting with light \citep{dicke,hepp,wang,emary,emary1}. If the atom-light interaction exceeds a critical value then the Dicke model undergoes a continuous phase transition to a state with non-vanishing photon number and discrete parity symmetry breaking. The Dicke model is reviewed with its applications in quantum optics in ref. \citep{garraway}. Furthermore, a direct implementation of a Dicke model Hamiltonian without any additional diamagnetic terms can also be provided in atomic experiments \citep{dimer}. Moreover, it was noted that a dynamical version of the superradiance transition in the Dicke model is equivalent to the self-organization transition \citep{baumann,baumann1,nagy}. The quantum phase transitions induced by the optical cavity field mediates long-range interactions among the atoms which may alter their behaviour. Recently, it has also been predicted that a cloud of atoms with extra transverse pump undergoes a self-organization transition to a spatially modulated phase \citep{domokos} which was experimentally proved using thermal clouds in an optical cavity \citep{black}. The phenomenon of self-organization has been recently observed in an experiment using a BEC in an optical cavity \citep{baumann,baumann1}. This experimental setup gives rise to a new kind of supersolid which is formed due to the spontaneous sublattice symmetry breaking coexisting with superfluid phase coherence \citep{andreev,chester,leggett}. It uses an optical cavity in which atomic states were replaced by the momentum states. Hence, the splitting between the states can be controlled by the atomic recoil energy. This enables the observation of Dicke model transition using light with optical frequencies.

Motivated by these interesting developments in the field of cavity optomechanics and ultracold gases, we propose an optomechanical system consisting of a two-level BEC within a high-finesse optomechanical cavity with one movable mirror. This system is used to study the theoretical analysis of the dissipative optomechanical Dicke model quantum phase transition within the thermodynamic limit. It is based on the linearized treatment of quantum fluctuations in the Holstein-Primakoff representation of the collective atomic spin and the standard input-output theory of open quantum-optical systems. We primarily study the effect of the mechanical mode of the oscillating mirror on the system dynamics with the variation in atom-photon coupling. The imaginary and real parts of the eigenvalues are analysed which demonstrates the quantum phase transition of the system. We further present the results for the cavity fluorescence spectrum that accounts for the quantum fluctuations and the homodyne spectra that measures the quadrature fluctuations of the cavity output field.

\section{The Basic Model}

The basic optomechanical system investigated here consists of a high-finesse Fabry-Perot optical cavity of length $L$ with one fixed mirror and another movable mirror of mass $m$, which is free to oscillate at some mechanical frequency $\omega_{m}$. Experimentally, a single vibrational mode of the movable mirror can be considered by using a bandpass filter in the detection scheme such that the other mechanical degrees of freedom arising from the radiation pressure can be neglected \citep{pinard}. In addition, our model involves an elongated cigar-shaped gas of $N$ two-level $^{87}Rb$ BEC atoms having mass $M$ and transition frequency $\omega_{a}$. The schematic representation of the system is illustrated in fig.(1). The BEC atoms are strongly coupled to a single one-dimensional quantized cavity mode of frequency $\omega_{c}$. The optical cavity is coherently driven by an external pump laser of frequency $\omega_{p}$ from a direction perpendicular to the cavity axis by acting as a constant source of photons for the cavity \citep{baumann}. For simplicity, we will consider the dynamics of the system along the axis of the cavity only. The radial motion of BEC is freezed out by a tight harmonic potential of frequency $\omega_{R}$ such that its spatial dimension along the cavity axis is taken into consideration only. The atom-pump detuning $\Delta_{a}(=\omega_{p}-\omega_{a})$ is assumed to be very large in order to suppress the spontaneous emission of photons by the atoms since this is a source of heat which can eventually destroy the condensate. Here, the electronically excited atomic state is adiabatically eliminated, which is justified for large atom-laser detuning. As a result, an effective two-level system is formed with two stable states: the atomic zero momentum state $|p>=|0>$ and the excited momentum state $|p>=|\pm \hbar k>$ which are coupled through a pair of distinct Raman channels \citep{baumann,dimer}. Here, $p$ denotes the momenta along the cavity axis and $k$ represents the wave vector of pump laser field. The effective atomic transition frequency $\omega_{0}$ is twice the atomic recoil frequency $\omega_{r}=\hbar k^{2}/2M$, namely $\omega_{0}=2\omega_{r}$. In our case, the measurement of the field quadratures of the cavity mode can be performed by homodyning the cavity output field using a local oscillator with an appropiate phase \citep{laurat,vitali1}.

The simplest model of the system, involving all the condensate atoms with different momentum states to be identically coupled with the single-mode cavity field, is provided by the following Hamiltonian in the dipole approximation \citep{bhattacherjee,dimer,baumann}:

\begin{equation}\label{ham1}
H=\hbar \omega_{c}c^{\dagger}c+\hbar \omega_{m}b^{\dagger}b+\hbar \omega_{0}S_{z}+\hbar \frac{g}{\sqrt{N}}(c+c^{\dagger})(S_{+}+S_{-})+\hbar \omega_{c} \eta_{0} c^{\dagger}c(b+b^{\dagger}).
\end{equation}

In Hamiltonian (\ref{ham1}), the first term describes the energy of the cavity mode where $c(c^{\dagger})$ is the annihilation (creation) operator of the cavity mode such that $[c,c^{\dagger}]=1$. The second term gives the energy of the single vibrational mode of the mechanical mirror with $b(b^{\dagger})$ as the annihilation (creation) operator such that $[b,b^{\dagger}]=1$. Third term represents the energy of the condensate atoms. Here, the ensemble of $N$ BEC atoms is described by using the picture of a collective spin which is basically the sum of effective spins $1/2$ that simply describe the internal degrees of freedom of each atom. The collective atomic operators are expressed as $S_{z}=\sum_{n}(|\pm \hbar k>_{n}$ $_{n}<\pm \hbar k|$ $-|0>_{n}$ $_{n}<0|)$ and  $S_{+}=S_{-}^{\dagger}=\sum_{n}|\pm \hbar k>_{n}$ $_{n}<0|$, with the index $n$ labelling the atoms. These operators satisfy the angular momentum commutation relations $[S_{+},S_{-}]=2S_{z}$ and $[S_{\pm},S_{z}]=\mp S_{\pm}$.

\begin{figure}[h]
\hspace{-0.0cm}
\includegraphics [scale=0.80]{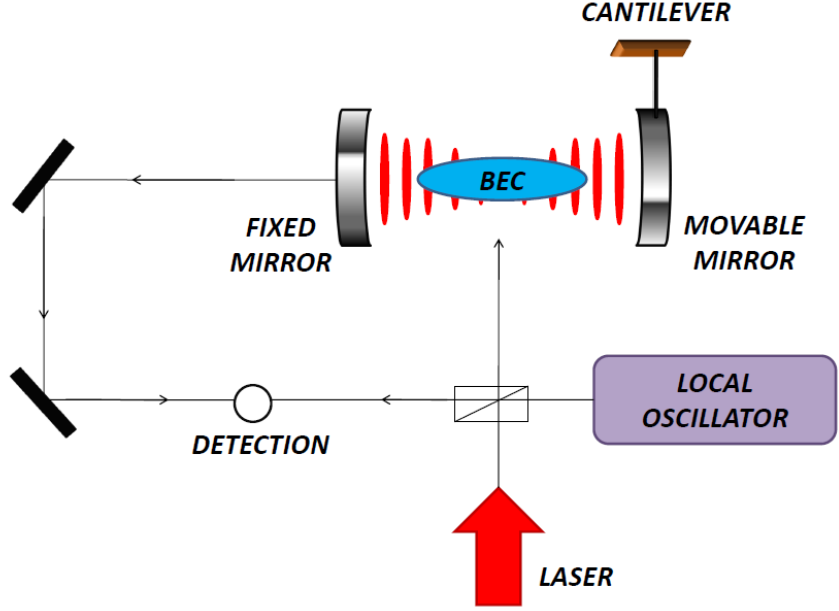}
\caption{(color online) Schematic representation of the setup. Figure shows an optomechanical system with Bose-Einstein Condensate confined in a high-finesse optical cavity driven by a transverse pump laser. A local oscillator is also provided using the beam splitter for the homodyne measurement of the light reflected by the cavity. Here one of the cavity mirrors is movable.}
\end{figure}\label{fig1}

Fourth term illustrates the interaction between the condensate field and the light field with $g$ being the collective atom-photon coupling strength, which can be tuned in experiment by varying the pump laser power \citep{baumann}. We are neglecting the contact interactions between the atoms of the condensate. Last term in the Hamiltonian represents the nonlinear dispersive coupling between the intensity of the cavity field and the position quadrature of the oscillating mirror which arises due to the radiation pressure force exerted by the intra-cavity photons on the movable mirror. This force exerted by the electromagnetic field on the movable mirror shifts the phase of the field by $2k_{0}L_{m}$, where $k_{0}$ is the propagation wave vector of the cavity field and $L_{m}$ denotes the displacement of the mirror from its equilibrium position. It basically depends upon the number of photons in the cavity. Here $\eta_{0}$ denotes the nonlinear dispersive coupling between the intensity of the cavity field and the position quadrature of the movable mirror with $\eta_{0}<<1$. For simplicity, we assume that the movable mirror has perfect reflectivity such that the light transmission from the cavity takes place through the fixed mirror only. Moreover, no direct coupling between the condensate atoms and cantilever is considered, which can be achieved by assuming that only a few lattice sites are appreciably populated near the centre of the cavity.

We now turn to discuss the dynamics arising from this model. Starting from the Hamitonian equation (\ref{ham1}), the coupled equations of motion for the operators $S_{z}$, $S_{-}$, $b$ and $c$ are given as:

\begin{equation}\label{eq1}
\dot{S_{z}}=i\frac{g}{\sqrt{N}}(c+c^{\dagger})(S_{-}-S_{+}),
\end{equation}

\begin{equation}\label{eq2}
\dot{S_{-}}=-i\omega_{0}S_{-}+2i\frac{g}{\sqrt{N}}(c+c^{\dagger})S_{z},
\end{equation}

\begin{equation}\label{eq3}
\dot{b}=-(\gamma_{m}+i\omega_{m})b-i\omega_{c}\eta_{0}c^{\dagger}c,
\end{equation}

\begin{equation}\label{eq4}
\dot{c}=-(\gamma_{c}+i\omega_{c})c-i\omega_{c}\eta_{0}c(b+b^{\dagger})-i\frac{g}{\sqrt{N}}(S_{+}+S_{-}),
\end{equation}

where $\gamma_{c}$ denotes the cavity loss rate which arises due to the leakage of photons through the mirrors. Also, the movable mirror is damped with decay constant $\gamma_{m}$ due to its interaction with the environment. The condensate atoms are robust, thus, we can neglect the effects of atom loss during the experimental time \citep{baumann}. In particular, the equations of motion (\ref{eq1})-(\ref{eq4}) are invariant under the parity transformation

\begin{equation}
S_{\pm}\rightarrow -S_{\pm}, c\rightarrow -c,
\end{equation}

as in the well-known Dicke model. This symmetry is spontaneously broken on passing from the normal phase to the superradiant phase \citep{emary1,baumann,baumann1}. Our starting point is the mean-field analysis of the system which can be demonstrated by introducing the c-number variables $Z\equiv<S_{z}>$, $\alpha\equiv<S_{-}>$, $\beta\equiv<b>$ and $\gamma\equiv<c>$, where $Z$, $\alpha$, $\beta$ and $\gamma$ are the population inversion, atomic polarization, mirror mode and complex cavity field amplitudes respectively. It has already been investigated using the coupled equations of motion (\ref{eq1})-(\ref{eq4}) in our previous work \citep{aggarwal}, which shows that the steady-state solution displays a bifurcation point at:

\begin{equation}
g=g_{c}\equiv \frac{1}{2}\sqrt{\frac{\omega_{0}(\gamma_{c}^{2}+\omega_{c}^{2})}{\omega_{c}}}.
\end{equation}

Here, $g_{c}$ denotes the critical value of atom-photon coupling strength. The steady-state solutions are evaluated with the constraint that the pseudo-angular momentum $W^{2}+\mid\alpha\mid^{2}=\frac{N^{2}}{4}$ is conserved. While $Z_{ss}=-N/2$ and $\alpha_{ss}=\beta_{ss}=\gamma_{ss}=0$ are the trivial steady state solutions for all values of $g$, they are only stable for $g<g_{c}$. For $g>g_{c}$, these solutions are no longer stable and new sets of stable steady-state solutions $Z_{ss}$, $\alpha_{ss}$, $\beta_{ss}$ and $\gamma_{ss}$ appear, which can be obtained from:

\begin{eqnarray}\label{eq5}
Z_{ss}^{3}\left[\frac{g^{2}\eta_{0}^{2}\Delta(1-2\epsilon_{1})}{N g_{c}^{2}}\right] +Z_{ss}\left[1-\frac{Ng^{2}\eta_{0}^{2}\Delta(1-2\epsilon_{1})}{4g_{c}^{2}} \right]+\frac{Ng_{c}^{2}}{2g^{2}}=0,
\end{eqnarray}

\begin{eqnarray}\label{eq6}
\alpha_{ss}=\pm\sqrt{\frac{N^{2}}{4}-Z_{ss}^{2}},
\end{eqnarray}

\begin{eqnarray}\label{eq7}
\beta_{ss}=\frac{-\omega_{c}\eta_{0}\mid\gamma_{ss}\mid^{2}(\omega_{m}+i\gamma_{m})}{\gamma_{m}^{2}+\omega_{m}^{2}},
\end{eqnarray}

\begin{eqnarray}\label{eq8}
\mid\gamma_{ss}\mid=\pm\left[\frac{N(\gamma_{c}^{2}+\omega_{c}^{2})}{4g^{2}\alpha_{ss}^{2}}-\frac{4\eta_{0}^{2}\omega_{m}\omega_{c}
\epsilon_{1}}{(\gamma_{m}^{2}+\omega_{m}^{2})}\right]^{-1/2},
\end{eqnarray}

\begin{eqnarray}\label{eq9}
\gamma_{ss}=\frac{-2ig\alpha_{ss}}{\sqrt{N}[\gamma_{c}+i\omega_{c}(1+2\eta_{0}Re[\beta_{ss}])]},
\end{eqnarray}

where $\epsilon_{1}=\frac{\omega_{c}^{2}}{\gamma_{c}^{2}+\omega_{c}^{2}}$ and $\Delta=\frac{2\omega_{m}\omega_{0}}{\gamma_{m}^{2}+\omega_{m}^{2}}$. Here $Z_{ss}$, $\alpha_{ss}$, $\beta_{ss}$, $\mid\gamma_{ss}\mid$ and $\gamma_{ss}$ are the steady state population inversion, polarization amplitude, mirror mode amplitude, absolute value of cavity-field amplitude and cavity-field amplitude respectively.

In the next section, we will discuss the Holstein-Primakoff representation to derive the effective Hamiltonian in the normal phase and in the superradiant phase within the thermodynamic limit.

\section{Thermodynamic limit}

In the thermodynamic limit of $N>>1$, the optomechanical Dicke Hamiltonian exhibits a quantum phase transition (QPT) at a critical value of the atom-cavity field coupling strength $g_{c}$, at which point the parity symmetry of the Hamiltonian is broken. In this analysis, we make an extensive use of the Holstein-Primakoff representation by expressing the atomic spin operators in terms of bosonic mode operators $a$ and $a^{\dagger}$ ($[a,a^{\dagger}]=1$) such that $S_{+}=a^{\dagger}(\sqrt{N-a^{\dagger}a})$, $S_{-}=S_{+}^{\dagger}=(\sqrt{N-a^{\dagger}a})a$ and $S_{z}=(a^{\dagger}a-\frac{N}{2})$ \citep{emary1,holstein}.

Making these subtitutions into the Hamiltonian of eqn.(\ref{ham1}), we obtain the following three-mode bosonic Hamitonian:

\begin{equation}\label{ham2}
H=\hbar \omega_{c}c^{\dagger}c+\hbar \omega_{m}b^{\dagger}b+\hbar \omega_{0}(a^{\dagger}a-\frac{N}{2})+\hbar g(c+c^{\dagger})\left[a^{\dagger}(\sqrt{1-\frac{a^{\dagger}a}{N}})+(\sqrt{1-\frac{a^{\dagger}a}{N}})a \right]+\hbar \omega_{c}\eta_{0}c^{\dagger}c(b+b^{\dagger}).
\end{equation}

The goal is to achieve the linearization about the semiclassical amplitudes mentioned in the previous section under the assumption $N>>1$. In the thermodynamic limit, quantum fluctuations are small and can be treated in a linearized approach.

In the normal phase ($g<g_{c}$), the semiclassical steady states $\alpha_{ss}$, $\beta_{ss}$ and $\gamma_{ss}$ are zero, thus, the expansion is made directly on the operators $a$, $b$ and $c$. This yields an effective Hamiltonian by simply neglecting terms with $N$ in the denominator in the full Hamiltonian (eqn.\ref{ham2}) which approximates the square root in the Holstein-Primakoff mapping with unity and is given as:

\begin{equation}\label{ham3}
H_{1}=\hbar \omega_{c}c^{\dagger}c+\hbar \omega_{m}b^{\dagger}b+\hbar \omega_{0}a^{\dagger}a+\hbar g(c+c^{\dagger})(a+a^{\dagger}).
\end{equation}

Additionally, to obtain the above Hamiltonian, we have omitted the constant terms and retained the terms that are bilinear in bosonic operators in the full Hamiltonian of eqn.(\ref{ham2}).

In the superradiant phase ($g>g_{c}$), the semiclassical steady states $\alpha_{ss}$, $\beta_{ss}$ and $\gamma_{ss}$ are nonzero and all the three bosonic modes $a$, $b$ and $c$ acquire macroscopic occupations. To do this, we start with the Holstein-Primakoff transformed Hamiltonian of eqn.(\ref{ham2}) and displace the bosonic modes in the following way:

\begin{equation}
a\rightarrow d+\frac{\alpha_{ss}}{\sqrt{\frac{N(1+\mu)}{2}}}, b\rightarrow e+\beta_{ss}, c\rightarrow f+\gamma_{ss}.
\end{equation}

Here $\alpha_{ss}$, $\beta_{ss}$ and $\gamma_{ss}$ are given in eqns.(\ref{eq6}), (\ref{eq7}) and (\ref{eq9}) respectively with $d$, $e$ and $f$ describing the quantum fluctuations around the semiclassical amplitudes. Moreover, we have defined $\mu=g_{c}^{2}/g^{2}<1$ . Making these transformations within the thermodynamic limit, we obtain the following Hamitonian (omitting constant terms):

\begin{equation}\label{ham4}
H_{2}=\hbar x_{1}d^{\dagger}d+\hbar \omega_{m}e^{\dagger}e+\hbar \omega_{1} f^{\dagger}f+\hbar x_{2}(d+d^{\dagger})(f+f^{\dagger})+\hbar x_{3}(e+e^{\dagger})(f+f^{\dagger})+\hbar x_{4}(d+d^{\dagger})^{2}+\hbar x_{5}(d+d^{\dagger})+\hbar x_{6}(e+e^{\dagger})+\hbar x_{7}(f+f^{\dagger}),
\end{equation}

where $\omega_{1}=\omega_{c}(1+2\eta_{0}\beta_{ss})$. The expressions for $x_{1}$, $x_{2}$, $x_{3}$, $x_{4}$, $x_{5}$, $x_{6}$ and $x_{7}$ are given in Appendix A. We now eliminate the terms in $H_{2}$ that are linear in bosonic operators by choosing $x_{5}=x_{6}=x_{7}=0$. This yields the final effective Hamitonian in the superradiant phase as:

\begin{equation}\label{ham5}
H_{2}=\hbar x_{1}d^{\dagger}d+\hbar \omega_{m}e^{\dagger}e+\hbar \omega_{1} f^{\dagger}f+\hbar x_{2}(d+d^{\dagger})(f+f^{\dagger})+\hbar x_{3}(e+e^{\dagger})(f+f^{\dagger})+\hbar x_{4}(d+d^{\dagger})^{2}.
\end{equation}

Having derived the two effective Hamiltonians that describe the system for all $g$ in the thermodynamic limit, we now study the system's properties in each of its two phases.

\section{Eigenvalue analysis}

In the normal phase ($g<g_{c}$), the effective bilinear Hamiltonian (\ref{ham3}) leads to the following coupled equations of motion for the expectation values of $a$, $b$ and $c$:

\begin{equation}
\dot{<a>}=-i\omega_{0}<a>-ig(<c>+<c^{\dagger}>),
\end{equation}

\begin{equation}
\dot{<b>}=-(\gamma_{m}+i\omega_{m})<b>,
\end{equation}

\begin{equation}
\dot{<c>}=-(\gamma_{c}+i\omega_{c})<c>-ig(<a>+<a^{\dagger}>).
\end{equation}

In matrix form, it can be written as $\dot{u_{1}}=M_{1}u_{1}$, where $u_{1}\equiv(<a>,<a^{\dagger}>,<b>,<b^{\dagger}>,<c>,<c^{\dagger}>)^{T}$ and $M_{1}$ is a constant $6\times6$ matrix given as:

\begin{equation}
         M_{1}=
            \left[ {\begin{array}{cccccc}
              -i\omega_{0} & 0 & 0 & 0 & -ig & -ig \\
              0 &  i\omega_{0} & 0 & 0 & ig & ig \\
              0 & 0 & (-i\omega_{m}-\gamma_{m}) & 0 & 0 & 0 \\
              0 & 0 & 0 & (i\omega_{m}-\gamma_{m}) & 0 & 0 \\
              -ig & -ig & 0 & 0 & (-i\omega_{c}-\gamma_{c}) & 0 \\
              ig &  ig & 0 & 0 & 0 & (i\omega_{c}-\gamma_{c}) \\
                \end{array} } \right].
\end{equation}

\begin{figure}[h]
\hspace{-0.0cm}
\begin{tabular}{cc}
\includegraphics [scale=0.80]{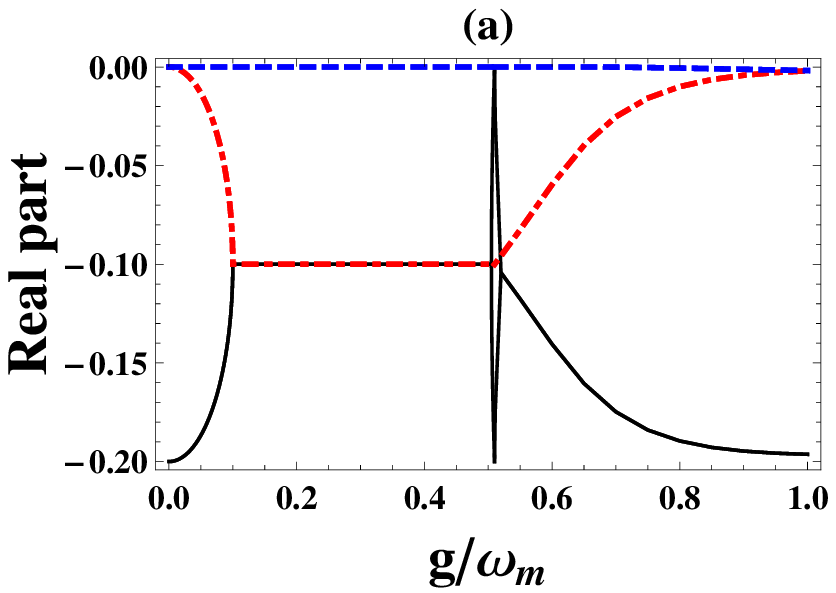}& \includegraphics [scale=0.80] {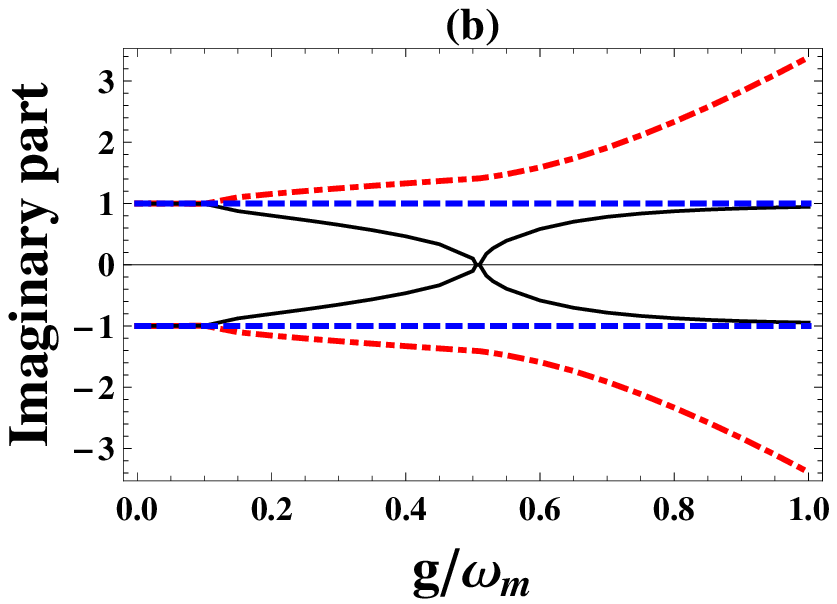}\\
\includegraphics [scale=0.80]{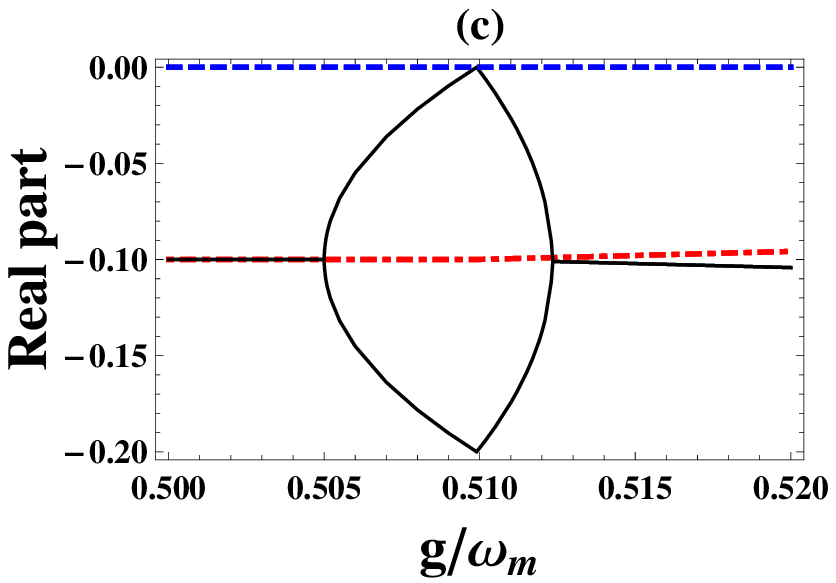}& \includegraphics [scale=0.80] {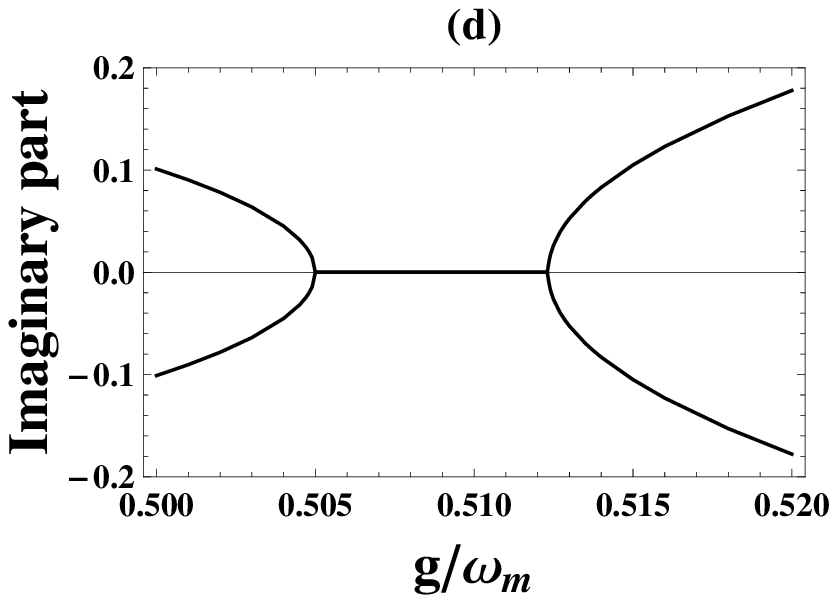}\\
\end{tabular}
\caption{(color online) Plot of the eigenvalues as a function of dimensionless atom-photon coupling strength ($g/\omega_{m}$). Figs.2(a) and 2(b) represent the real and imaginary parts of the eigenvalues respectively as a function of $g$. Figs.2(c) and 2(d) show the magnified views of respective real and imaginary parts around $g=g_{c}=0.5099 \omega_{m}$. Solid lines associate with the photonic branch, dot dashed lines with the atomic branch and dashed lines with the phononic branch. The parameters used are $\omega_{c}=\omega_{0}=\omega_{m}$, $\gamma_{c}=0.2\omega_{m}$, $\gamma_{m}=10^{-5}\omega_{m}$, $N=10$ and $\eta_{0}=0.01$.}
\end{figure}\label{fig2}

Similarly, in the superradiant phase ($g>g_{c}$), the coupled equations of motion for the expectation values of $d$, $e$ and $f$, obtained using the effective quadratic Hamiltonian (\ref{ham5}), are:

\begin{equation}
\dot{<d>}=-ix_{1}<d>-ix_{2}(<f>+<f^{\dagger}>)-2ix_{4}(<d>+<d^{\dagger}>),
\end{equation}

\begin{equation}
\dot{<e>}=-(i\omega_{m}+\gamma_{m})<e>-ix_{3}(<f>+<f^{\dagger}>),
\end{equation}

\begin{equation}
\dot{<f>}=-(i\omega_{1} +\gamma_{c}) <f>-ix_{2}(<d>+<d^{\dagger}>)-ix_{3}(<e>+<e^{\dagger}>).
\end{equation}

Thus, we can write it in a matrix form as $\dot{u_{2}}=M_{2}u_{2}$ with $u_{2}\equiv(<d>,<d^{\dagger}>,<e>,<e^{\dagger}>,<f>,<f^{\dagger}>)^{T}$ and the constant $6\times6$ matrix $M_{2}$ becomes:

\begin{equation}
         M_{2}=
            \left[ {\begin{array}{cccccc}
              -i(x_{1}+2x_{4}) & -2ix_{4} & 0 & 0 & -ix_{2} & -ix_{2} \\
              2ix_{4} &  i(x_{1}+2x_{4}) & 0 & 0 & ix_{2} & ix_{2} \\
              0 & 0 & (-i\omega_{m}-\gamma_{m}) & 0 & -ix_{3} & -ix_{3} \\
              0 & 0 & 0 & (i\omega_{m}-\gamma_{m}) & ix_{3} & ix_{3} \\
              -ix_{2} & -ix_{2} & -ix_{3} & -ix_{3} & (-i\omega_{1}-\gamma_{c}) & 0 \\
             ix_{2} & ix_{2} & ix_{3} & ix_{3} & 0 & (i\omega_{1}-\gamma_{c}) \\
                \end{array} } \right].
\end{equation}

The eigenvalues of $M_{1}$ and $M_{2}$ are evaluated with the help of MATHEMATICA 9.0. They are plotted as a function of dimensionless atom-cavity field coupling strength ($g/\omega_{m}$) in fig.2 for $\omega_{c}=\omega_{m}=\omega_{0}$, with the six eigenvalues grouped into pairs. One of the pairs is associated with the atomic branch (dot dashed lines), another with the photonic branch (solid lines) and the remaining one with the phononic branch (dashed lines). Figs.2(a) and 2(b) represent the real part and the imaginary part of the eigenvalues in the linearized Holstein-Primakoff representation respectively. Figs.2(c) and 2(d) depict the magnified views of the respective real and imaginary parts around the quantum phase transition at $g=g_{c}$. The phononic branch eigenvalues take on the constant and imaginary values $\pm i\omega_{m}$ with almost zero real parts for all values of $g$ (see figs.2(a) and 2(b)). In addition to $g_{c}$, there are two other significant atom-photon coupling strengths $g_{1}=0.5050 \omega_{m}$ and $g_{2}=0.5124 \omega_{m}$ for the dispersive cavity case (non-zero cavity decay rate). Fig.2(c) illustrates that, as $g$ approaches $g_{1}$, the real parts of the eigenvalues associated with the photonic branch split such that the real part of one of the eigenvalues go to zero at the phase transition point $g_{c}$. However, the imaginary parts of the corresponding eigenvalues become zero as $g \rightarrow g_{1}$ and remain zero in the interval $g_{1}<g<g_{2}$ (see fig.2(d)). Further note that for $g>g_{2}$, the eigenvalues on the photonic branch take on the nonzero imaginary parts once again. For large $g$,  photonic branch eigenvalues approach the value $-\gamma_{c} \pm i\omega_{c}$. In the case of atomic branch, fig.2(a) shows that the real parts of the eigenvalues above the critical point decrease with increase in $g$ and approach zero value for large $g$. In correspondence, the imaginary parts of the eigenvalues move away from the $\pm i \omega_{0}$ value with increase in $g$ (see fig.2(b)). Moreover, no splitting is observed in the real parts of the atomic branch eigenvalues for all the values of atom-light coupling, which can be seen from figs.2(a) and 2(c).

Thus, the eigenvalue analysis provides a first outlook of the quantum fluctuations with the atom-light field coupling which we will see in detail in the next sections by monitoring the photons that leak out of the cavity.

\section{Input-Output Theory}

In the previous sections, the equations of motion describe the internal dynamics of the system. From now onwards, we consider measurements on the light leaving out the system through the cavity output mirror in order to probe this dynamics. To this end, we make use of the standard input-output formalism \citep{collett,gardiner,walls} by introducing the cavity input and output field noise operators $c_{in}(t)$ and $c_{out}(t)$ such that: for $g<g_{c}$

\begin{equation}\label{eq10}
c_{out}(t)=(\sqrt{2\gamma_{c}})c(t)-c_{in}(t),
\end{equation}

and for $g>g_{c}$,

\begin{equation}\label{eq10a}
c_{out}(t)=(\sqrt{2\gamma_{c}})(f(t)+\gamma_{ss})-c_{in}(t).
\end{equation}

The correlation functions for the input noise operators are given in Appendix B. With the help of these cavity input field correlations, the cavity output field correlation functions can be calculated from eqns.(\ref{eq10}) and (\ref{eq10a}). The quantum Langevin equations of the system by considering position and momentum quadratures of the oscillating mirror, defined as $q_{1}(t)=[b(t)+b^{\dagger}(t)]$ and $p_{1}(t)=i[b^{\dagger}(t)-b(t)]$ respectively, for $g<g_{c}$ are:

\begin{equation}
\dot{a}(t)=-i\omega_{0}a(t)-ig(c(t)+c^{\dagger}(t)),
\end{equation}

\begin{equation}
\dot{c}(t)=-(\gamma_{c}+i\omega_{c})c(t)-ig(a(t)+a^{\dagger}(t))+\sqrt{2\gamma_{c}}c_{in}(t),
\end{equation}

\begin{equation}
\dot{q_{1}}(t)=\omega_{m}p_{1}(t),
\end{equation}

\begin{equation}
\dot{p_{1}}(t)=-\omega_{m}q_{1}(t)-\gamma_{m}p_{1}(t)+W(t).
\end{equation}

Here $W(t)=i\sqrt{\gamma_{m}}[\xi_{m}^{\dagger}(t)-\xi_{m}(t)]$ satisfies the correlation given in Appendix B, with $\xi_{m}(t)$ representing the Brownian noise operator arising due to the mechanical motion of the movable mirror.

Similarly, for $g>g_{c}$, the quantum Langevin equations are given as:

\begin{equation}
\dot{d}(t)=-ix_{1}d(t)-ix_{2}(f(t)+f^{\dagger}(t))-2ix_{4}(d(t)+d^{\dagger}(t)),
\end{equation}

\begin{equation}
\dot{f}(t)=-(i\omega_{1} +\gamma_{c})f(t)-ix_{2}(d(t)+d^{\dagger}(t))-ix_{3}(e(t)+e^{\dagger}(t))+\sqrt{2\gamma_{c}}c_{in}(t),
\end{equation}

\begin{equation}
\dot{q_{2}}(t)=\omega_{m}p_{2}(t),
\end{equation}

\begin{equation}
\dot{p_{2}}(t)=-\omega_{m}q_{2}(t)-\gamma_{m}p_{2}(t)-2x_{3}(f(t)+f^{\dagger}(t))+W(t).
\end{equation}

where $q_{2}(t)=[e(t)+e^{\dagger}(t)]$ and $p_{2}(t)=i[e^{\dagger}(t)-e(t)]$ are the respective postion and momentum quadratures of the movable mirror. Now, we compute the quantum Langevin equations in the Fourier space by using the following definitions of Fourier transforms:

\begin{equation}
F(\omega)=\frac{1}{\sqrt{2\pi}}\int_{-\infty}^\infty e^{i\omega t} F(t)dt,
\end{equation}

\begin{equation}
F^{\dagger}(-\omega)=\frac{1}{\sqrt{2\pi}}\int_{-\infty}^\infty e^{i\omega t} F^{\dagger}(t)dt,
\end{equation}

where, $F$ represents any of the operators $a$, $b$, $c$, $d$, $e$, $f$, $c_{in}$ or $W$. Thus, the system operators in the frequency space take the following form: for $g<g_{c}$,

\begin{equation}
a(\omega)=\frac{-g}{(\omega_{0}-\omega)}[c(\omega)+c^{\dagger}(-\omega)],
\end{equation}

\begin{equation}\label{eq11}
c(\omega)=\frac{\sqrt{2\gamma_{c}}\left\lbrace \left[\left\lbrace \gamma_{c}-i(\omega+\omega_{c}) \right\rbrace(\omega^{2}-\omega_{0}^{2})-2i\omega_{0}g^{2}\right] c_{in}(\omega)-2i\omega_{0}g^{2}c_{in}^{\dagger}(-\omega)\right\rbrace }{\left\lbrace \gamma_{c}-i(\omega+\omega_{c})\right\rbrace \left\lbrace \gamma_{c}-i(\omega -\omega_{c}) \right\rbrace(\omega^{2}-\omega_{0}^{2})+4\omega_{0}^{2}g^{2}} ,
\end{equation}

\begin{equation}
q_{1}(\omega)=\frac{-\omega_{m}W(\omega)}{(\omega^{2}-\omega_{m}^{2}+i\omega \gamma_{m})},
\end{equation}

\begin{equation}
p_{1}(\omega)=\frac{\omega_{m}q_{1}(\omega)-W(\omega)}{(i\omega -\gamma_{m})},
\end{equation}

whereas for $g>g_{c}$,

\begin{equation}
d(\omega)=\frac{-x_{2}(x_{1}+\omega)\left[f(\omega)+f^{\dagger}(-\omega) \right]}{(x_{1}^{2}+4x_{1}x_{4}-\omega^{2})},
\end{equation}

\begin{eqnarray}\label{eq12}
f(\omega)=\frac{\left[W(\omega)\left\lbrace 2A_{4}(\omega)A_{3}(\omega)+iA_{4}(\omega)A_{2}(\omega)\right\rbrace +\sqrt{2\gamma_{c}}A_{5}(\omega)A_{2}(\omega)c_{in}(\omega)+2i\sqrt{2\gamma_{c}}A_{5}(\omega)A_{3}(\omega)c_{in}^{\dagger}(-\omega) \right]}{(A_{1}(\omega)A_{2}(\omega)-4A_{3}^{2}(\omega))},
\end{eqnarray}

\begin{equation}
q_{2}(\omega)=\frac{\omega_{m}}{(\omega^{2}-\omega_{m}^{2}+i\omega \gamma_{m})}\left[2x_{3}\left\lbrace f(\omega)+f^{\dagger}(-\omega)\right\rbrace -W(\omega) \right],
\end{equation}

\begin{equation}
p_{2}(\omega)=\frac{\omega_{m}q_{2}(\omega)+2x_{3}\left\lbrace f(\omega)+f^{\dagger}(-\omega)\right\rbrace -W(\omega)}{(i\omega -\gamma_{m})}.
\end{equation}

The expressions for $A_{1}(\omega)$, $A_{2}(\omega)$, $A_{3}(\omega)$, $A_{4}(\omega)$ and $A_{5}(\omega)$ are given in Appendix C. In the next section, we study the fluorescence spectrum of the cavity output field in each of the two phases, namely, the normal phase and the superradiant phase.

\section{Fluorescence Spectrum}

Fluorescence spectrum (or power spectrum) is proportional to the probability of detecting a photon of frequency $\omega$ at the cavity output. It basically consists of an incoherent part that accounts for quantum fluctuations, a coherent part that represents the mean excitation of the intracavity field and the semiclassical steady state amplitude $\gamma_{ss}$. The photon flux measured outside the cavity can be expressed as $<c_{out}^{\dagger}(t)c_{out}(t)>$. Thus, the incoherent part of the cavity fluorescence spectrum can be defined as:

\begin{equation}
S(\omega)\delta(\omega+\omega')=<c_{out}^{\dagger}(-\omega), c_{out}(\omega')>.
\end{equation}

It can be computed using the solutions to the quantum Langevin equations (\ref{eq11}) and (\ref{eq12}) and the input-output relations

\begin{equation}\label{12a}
c_{out}(\omega)=(\sqrt{2\gamma_{c}})c(\omega)-c_{in}(\omega),
\end{equation}

$g<g_{c}$, and

\begin{equation}\label{12b}
c_{out}(\omega)=(\sqrt{2\gamma_{c}})[f(\omega)+\sqrt{2\pi}\gamma_{ss}\delta(\omega)]-c_{in}(\omega),
\end{equation}

$g>g_{c}$. Thus, using the correlations given in Appendix B, the incoherent part of the cavity Fluorescence spectrum in the normal phase ($g<g_{c}$) becomes:

\begin{equation}\label{eq13}
S^{(1)}(\omega)=\frac{32\pi \gamma_{c}^{2}g^{4}\omega_{0}^{2}}{B_{1}(\omega)B_{1}^{\dagger}(\omega)}.
\end{equation}

\begin{figure}[h]
\hspace{-0.0cm}
\begin{tabular}{cc}
\includegraphics [scale=0.80]{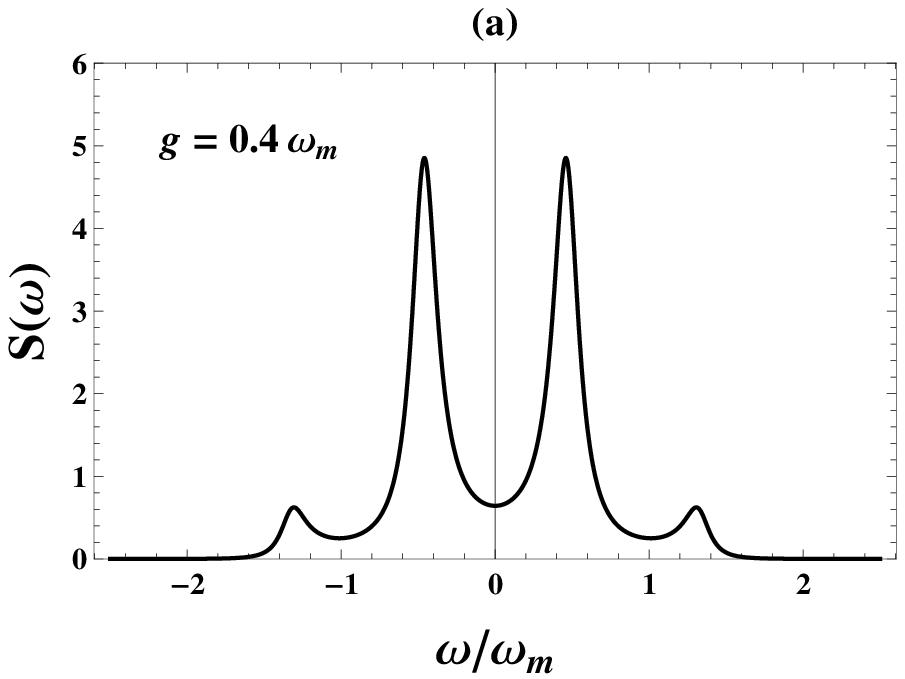}& \includegraphics [scale=0.80] {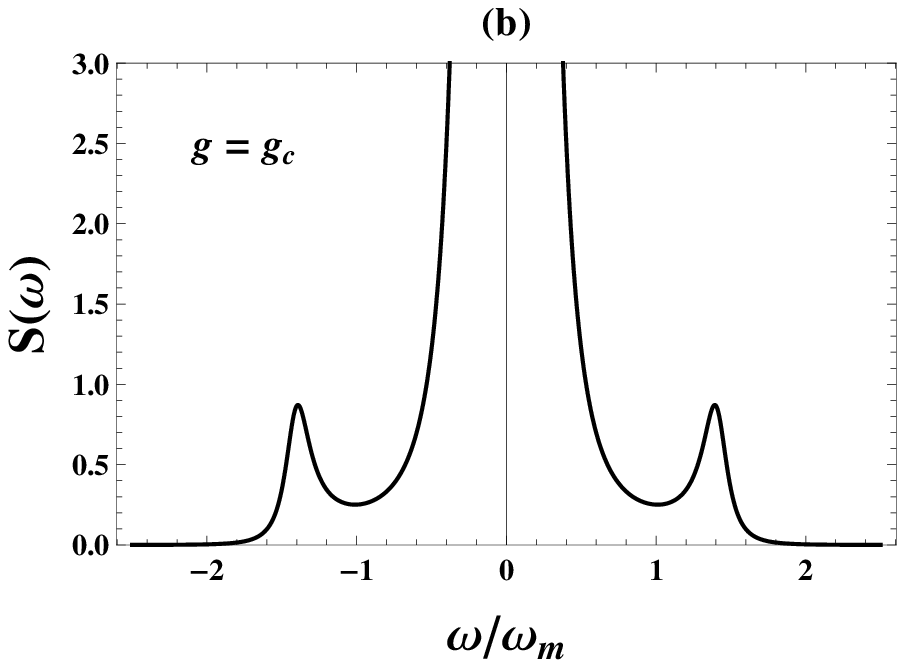}\\
\includegraphics [scale=0.80]{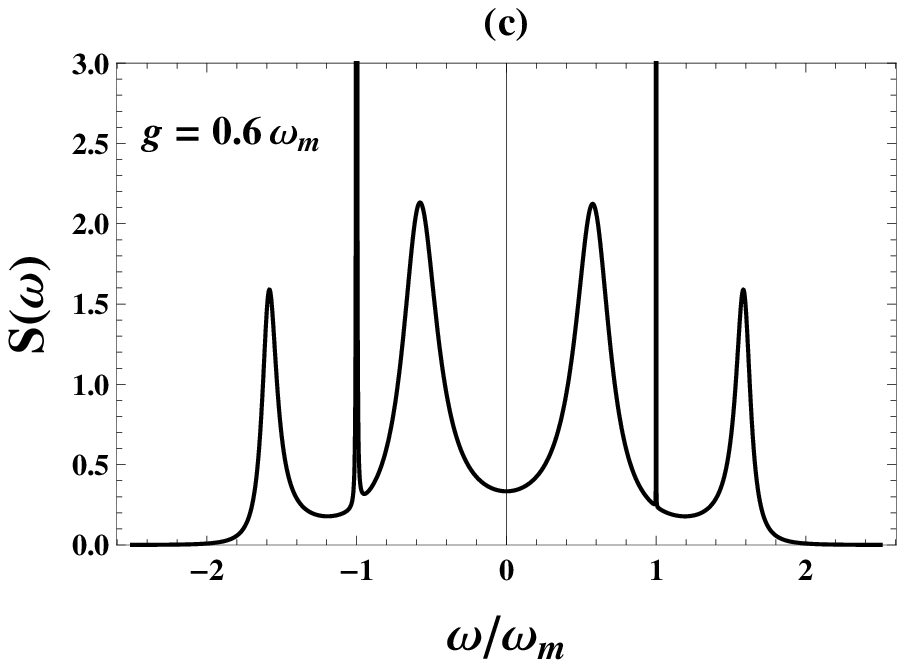}& \includegraphics [scale=0.80] {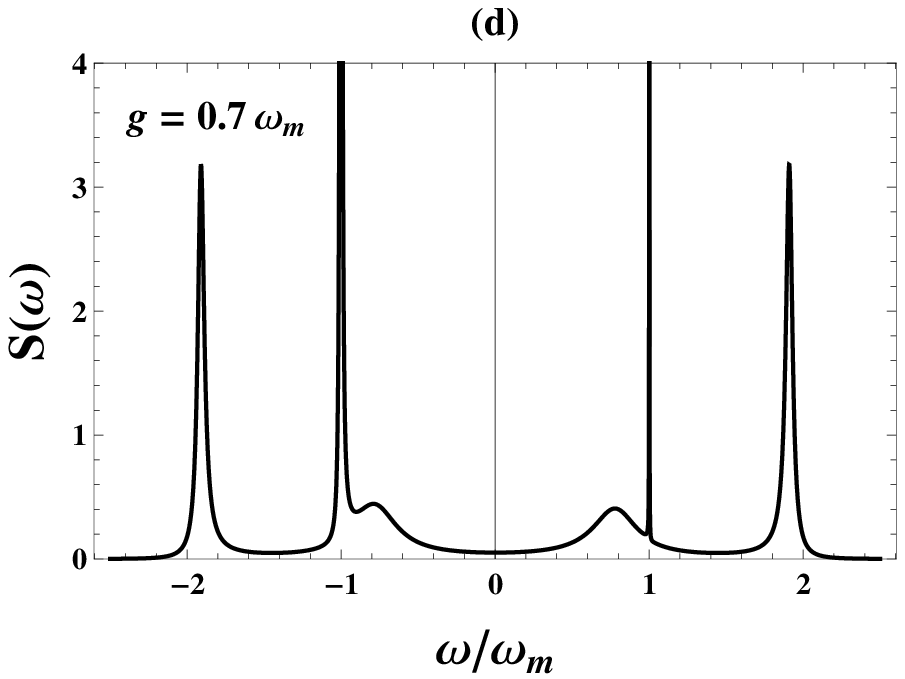}\\
\end{tabular}
\caption{(color online) Plot of incoherent part of fluorescence spectrum $S(\omega)$ for different values of atom-cavity field coupling strength $g=0.4\omega_{m}$ (plot (a)), $g=g_{c}=0.5099 \omega_{m}$ (plot (b)), $g=0.6\omega_{m}$ (plot(c)) and $g=0.7\omega_{m}$ (plot (d)). The other parameters used are $\omega_{c}=\omega_{0}=\omega_{m}$, $\gamma_{c}=0.2\omega_{m}$, $\gamma_{m}=10^{-5}\omega_{m}$, $N=10$, $k_{B}T/\hbar\omega_{m}=10^{4}$ and $\eta_{0}=0.01$.}
\end{figure}\label{fig3}

Similarly, in the superradiant phase ($g>g_{c}$), it is given as:

\begin{equation}\label{eq14}
S^{(2)}(\omega)=\frac{2\gamma_{c}[B_{2}(\omega)+B_{3}(\omega)+B_{4}(\omega)+B_{5}(\omega)+B_{6}(\omega)]}{\left[A_{1}(-\omega)A_{2}(-\omega)-4(A_{3}(-\omega))^{2} \right]\left[A_{1}^{\dagger}(-\omega)A_{2}^{\dagger}(-\omega)-4(A_{3}^{\dagger}(-\omega))^{2} \right] }.
\end{equation}

The values of $B_{1}(\omega)$, $B_{2}(\omega)$, $B_{3}(\omega)$, $B_{4}(\omega)$, $B_{5}(\omega)$ and $B_{6}(\omega)$ are mentioned in Appendix C. The incoherent part of fluorescence spectrum is plotted in fig.3 for four different values of atom-photon coupling strength $g=0.4\omega_{m}$ (plot (a)), $g=g_{c}$ (plot (b)), $g=0.6\omega_{m}$ (plot(c)) and $g=0.7\omega_{m}$ (plot (d)). For a particular value of $g$, the positions and widths of the spectral peaks are determined by the eigenvalues discussed in section IV, which can be clearly seen from the figs.2 and 3. Below the transition point, the central and outer doublets are observed which are associated with the photonic and atomic branch eigenvalues respectively (see fig.3(a)). As $g \rightarrow g_{c}$, fig.3(b) illustrates that the spectral peaks corresponding to the photonic branch doublet merge and forms a single narrow peak at $\omega=0$. The intensity under this peak diverges at $g=g_{c}$. Above the transition point, a pair of photonic branch doublets appears again (see figs.3(c) and 3(d)). In this case, the spectrum exhibits another pair of doublets associated with the phononic branch appearing in between the photonic and atomic doublets. The presence of the three pair of doublets in the fluoresence spectra of the cavity output field above the critical point is due to the coupling between the cavity field fluctuations, condensate fluctuations (Bogoliubov mode) and the mechanical mode fluctuations. This coupling between the three modes leads to the splitting of normal mode into three modes (normal mode splitting (NMS)) on both the positive frequency side and the negative frequency side, which is clearly indicated by the presence of six spectral peaks in figs.3(c) and 3(d). Normal mode basically refers to the mode characterizing small deviation of the field from its steady state. NMS involves driving three parametrically coupled nondegenerate modes out of equilibrium which further indicates the coherent energy exchange between the mechanical mode, cavity mode and the Bogoliubov mode. This energy exchange should take place on a time scale faster than the decoherence of each mode. Another vital observation apparent from figs. 3(c) and 3(d) is the asymmetric coherent energy exchange between the three bosonic modes (mechanical mode, cavity mode and the Bogoliubov mode) with the change in sign of $\omega$ in the superradiant phase ($g>g_{c}$). As the atom-photon coupling increases, the effect of counter-rotating components for the atomic system becomes perceptible. When the counter-rotating components are taken into consideration, a driving field resonant with the two-level transition is no longer resonant due to Bloch-Siegert effect \citep{bloch,zhang,chang}. Moreover, the motion of mechanical resonator via radiation pressure also changes the cavity's resonance frequency. Such an off resonant field results in asymmetric Autler-Townes splitting in the presence of optomechanical coupling, which can be clearly seen in figs.3(c) and 3(d). This asymmetry increases with increase in atom-cavity coupling. This is the most important result that we have observed in this paper. However, in the absence of optomechanical interaction, symmetric Autler-Townes splitting appears for all values of $g$ \citep{dimer}.

Figs.3(c) and 3(d) further show that the positions of the phononic branch peaks do not change with the increase in atom-cavity coupling. It is because of the fact that the eigenvalues associated with the phononic branch remains constant at $\pm i \omega_{m}$ for all values of $g$. Moreover, it is observed that the atomic branch peaks move linearly apart and become increasingly sharp with the increase in atom-cavity coupling. Far above the critical point, the photonic branch peaks approach the cavity mode resonance frequency $\omega=\pm \omega_{c}=\pm \omega_{m}$. Also note that, below the critical point, the nonexistence of the peaks of the phononic branch doublet is due to the absence of the optomechanical coupling in the normal phase. In the next section, we investigate the homodyne spectrum of the cavity output field in both the phases- normal phase and superradiant phase.

\section{Homodyne Spectrum}

Homodyne spectrum basically measures the quadrature noise (fluctuation) variances in the Fourier space of the output field quadrature amplitudes. The output field quadrature operator in time space can be given as:

\begin{equation}
Q_{out,\theta}(t)=\frac{\left[c_{out}(t)e^{-i\theta}+c_{out}^{\dagger}(t)e^{i\theta}\right] }{2},
\end{equation}

where $\theta$ represents the quadrature phase. In the Fourier space, it becomes:

\begin{equation}
Q_{out,\theta}(\omega)=\frac{\left[c_{out}(\omega)e^{-i\theta}+c_{out}^{\dagger}(-\omega)e^{i\theta}\right] }{2}.
\end{equation}

The homodyne spectrum for the output cavity field ($S_{out,\theta}(\omega)$) in the normally ordered form can be defined as \citep{collett,walls}:

\begin{equation}
S_{out,\theta}(\omega)\delta(\omega +\omega')=<Q_{out,\theta}(\omega),Q_{out,\theta}(\omega')>,
\end{equation}

which can be evaluated by using the solutions for the intracavity fields (\ref{eq11}) and (\ref{eq12}) and the input-output relations (\ref{12a}) and (\ref{12b}).

\begin{figure}[h]
\hspace{-0.0cm}
\begin{tabular}{cc}
\includegraphics [scale=0.80]{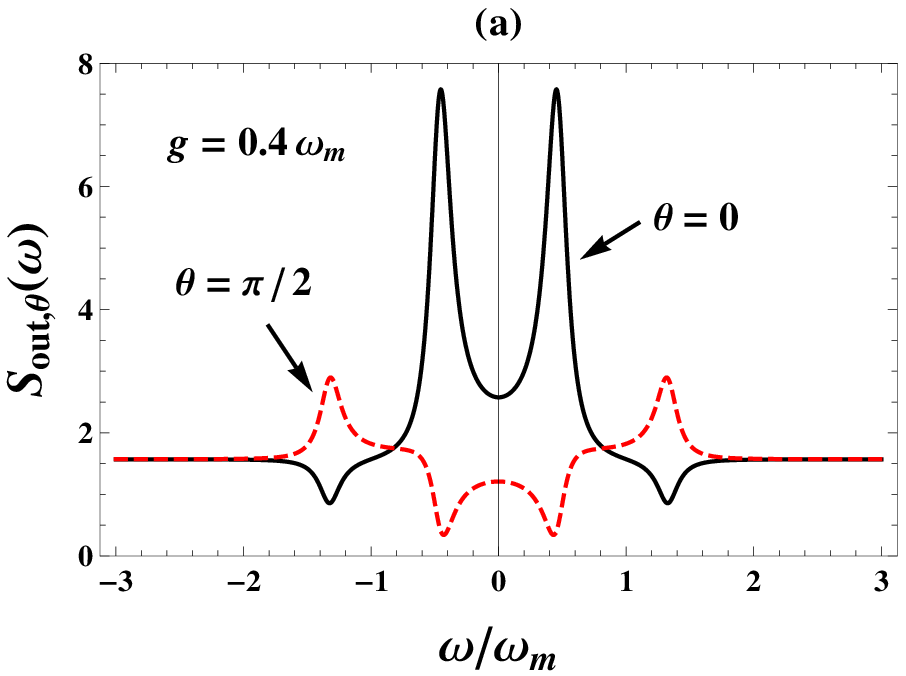}& \includegraphics [scale=0.80] {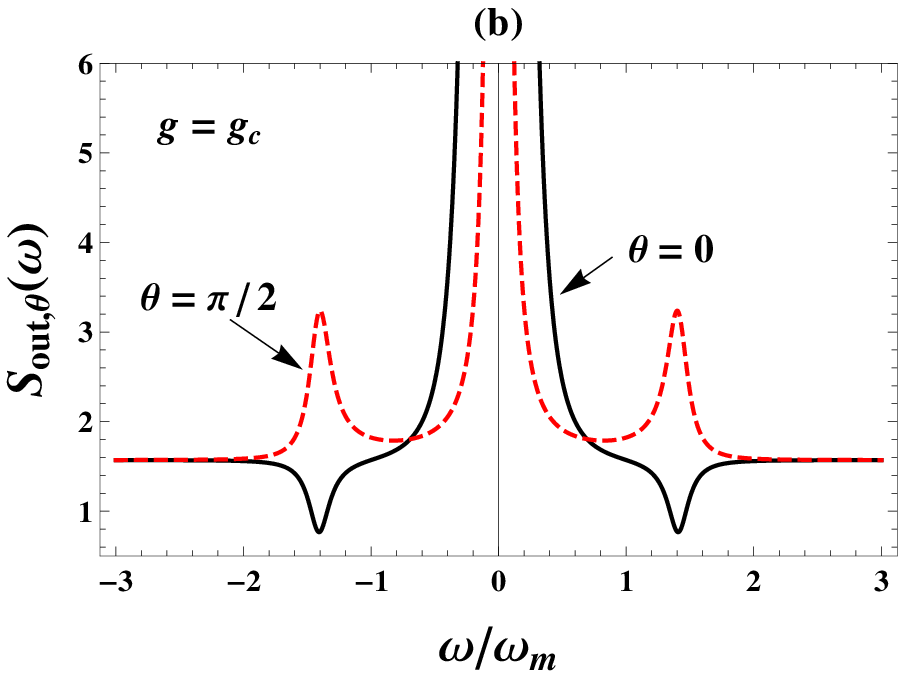}\\
\includegraphics [scale=0.80]{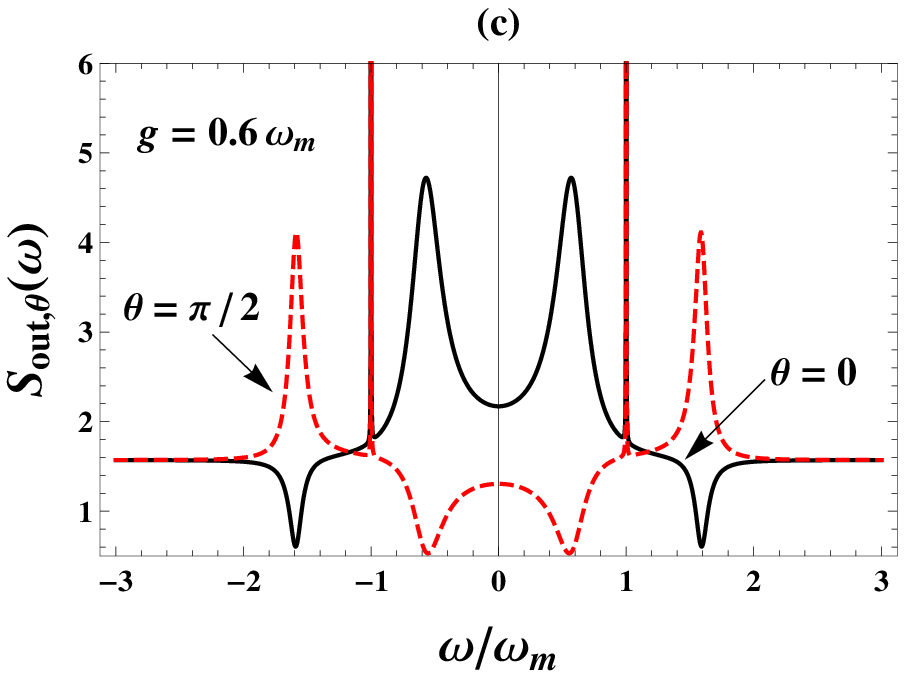}& \includegraphics [scale=0.80] {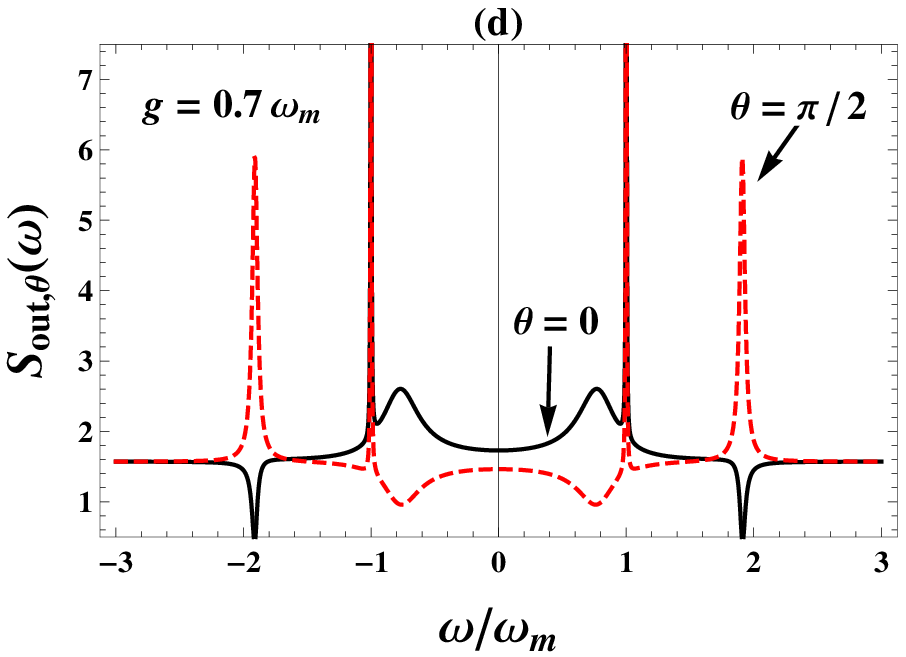}\\
\end{tabular}
\caption{(color online) Plot of homodyne spectrum $S_{out,\theta}(\omega)$ for different values of atom-photon coupling strength $g=0.4\omega_{m}$ (plot (a)), $g=g_{c}=0.5099 \omega_{m}$ (plot (b)), $g=0.6\omega_{m}$ (plot(c)) and $g=0.7\omega_{m}$ (plot (d)) with $\theta=0$ (solid line) and $\theta=\pi/2$ (dashed line). The other parameters used are $\omega_{c}=\omega_{0}=\omega_{m}=1$, $\gamma_{c}=0.2\omega_{m}$, $\gamma_{m}=10^{-5}\omega_{m}$, $k_{B}T/\hbar \omega_{m}=10^{4}$, $N=10$ and $\eta_{0}=0.01$.}
\end{figure}\label{fig4}

Hence, with the help of correlations given in Appendix B, the homodyne spectra in the normal phase becomes:

\begin{equation}\label{eq15}
S_{out,\theta}^{(1)}(\omega)=\frac{e^{-2i\theta}C_{1}(\omega)+C_{2}(\omega)+C_{3}(\omega)+e^{2i\theta}C_{4}(\omega)}{4}.
\end{equation}

Similarly, the expression for the homodyne spectra in the superradiant phase becomes:

\begin{equation}\label{eq16}
S_{out,\theta}^{(2)}(\omega)=\frac{e^{-2i\theta}D_{1}(\omega)+D_{2}(\omega)+D_{3}(\omega)+e^{2i\theta}D_{4}(\omega)}{4}.
\end{equation}

The expressions for $C_{1}(\omega)$, $C_{2}(\omega)$, $C_{3}(\omega)$, $C_{4}(\omega)$, $D_{1}(\omega)$, $D_{2}(\omega)$, $D_{3}(\omega)$ and $D_{4}(\omega)$ are illustrated in Appendix D. Fig.4 represents the homodyne spectrum (or quadrature noise spectrum) $S_{out,\theta}(\omega)$ for four different values of atom-photon coupling strength $g=0.4\omega_{m}$ (plot (a)), $g=g_{c}$ (plot (b)), $g=0.6\omega_{m}$ (plot(c)) and $g=0.7\omega_{m}$ (plot (d)) with $\theta=0$ (solid line) and $\theta=\pi/2$ (dashed line). Here also, we have chosen the same set of values of atom-photon coupling strength to correspond to fig.3. The numerical results for the quadrature phase $\theta=0$ and $\theta=\pi/2$ are well presented in the figure. For $\theta=0$, some of the observations in the homodyne spectra as a function of $g$ are same as displayed in the fluorescence spectra. The lower frequency peaks are associated with the photonic branch, the higher frequency peaks with the atomic branch and the intermediate frquency peaks with the phononic branch. As the atom-photon coupling strength approaches its critical value $g_{c}$, the phase transition is signaled by a divergence of the quadrature amplitude fluctuations at $\omega=0$ for both of the quadrature phases $\theta=0$ and $\theta=\pi/2$ (shown in fig.4(b)). Further note that the photonic and atomic spectral peaks in the noise spectrum get inverted at $\theta+\pi/2$ quadrature phase. However, the spectral peaks corresponding to phononic branch overlap for both of the quadrature phases $\theta=0$ and $\theta=\pi/2$ . Thus, the quantum noise variances or the fluctuation variances of the output field quadrature amplitudes can be effectively measured with the help of homodyne spectra. In principle, the homodyne spectra of the output cavity field can be used to determine the variance-based measures of atom-cavity field entanglement, which has been shown briefly in a recent paper \citep{dimer}. Further note that the homodyne spectra is completely symmetric. This symmetry of the spectra is ensured by the energy conservation such that the values of $S_{out,\theta}^{(1)}(\omega)$ and $S_{out,\theta}^{(2)}(\omega)$ remain unaffected with the change in sign of $\omega$.

Now, the experimentally realizable parameters used in the main paper to demonstrate the dynamics of the system are illustrated as follows. The mechanical frequency of the mirror in an optomechanical system can be varied from $2\pi \times 100$Hz \citep{cole}, $2\pi \times 10$kHz \citep{hunger}, to $2\pi \times 73.5$MHz \citep{schliesser1} with the corresponding damping rate from $2\pi \times 10^{-3}$Hz \citep{cole}, $2\pi \times 3.22$Hz \citep{hunger}, to $2\pi \times 1.3$kHz \citep{schliesser1}. The cavity field can have damping rate $2\pi \times 1.3$MHz \citep{brennecke} ($2\pi \times 0.66$MHz \citep{murch}). A cloud of BEC interacting with the light field of a high-finesse Fabry-Perot cavity may have a coherent coupling strength of $2\pi \times 10.9$MHz \citep{brennecke} ($2\pi \times 14.4$MHz \citep{murch}), which is significantly larger than the cavity decay rate. It, thus, places the system firmly in a regime where the Hamiltonian dynamics dominate. The high-finesse optical cavity is used to minimize the loss of photons through the cavity mirrors in order to have strong atom-cavity field coupling. The mirror-photon coupling rate is $2\pi \times 2.0$MHz. In fact $\frac{k_{B}T}{\hbar}\simeq 10^{11}$ $s^{-1}$ even at cryogenic temperatures, thus, it is always much larger than all the other parameters. Hence, even for high values of $\omega$, one can safely approximate $(\gamma_{m}\omega/\omega_{m})\left\lbrace 1+\coth[(\hbar \omega)/(2k_{B}T)]\right\rbrace \simeq (2\gamma_{m}k_{B}T)/(\hbar\omega_{m})$. In typical optomechanical experiments, the limit $\hbar \gamma_{m}<<\hbar\omega_{m}<<k_{B}T$ is always taken into account \citep{hadjar,tittonen,cohadon,pinard1}. Also, in a regime of strong-coupling cavity quantum electrodynamics, the critical regime of the Dicke model can be realized with just a few atoms \citep{mckeever,sauer,maunz}.

\section{Conclusion}

In conclusion, we have analyzed a dissipative optomechanical Dicke model in the thermodynamic limit for the detailed study of quantum phase transition involving a collective atomic pseudospin, a single quantized mode of the electromagnetic field and a single quantized mechanical mode of the movable mirror. In the eigenvalue analysis of the system, six eigenvalues have been obtained which are grouped into three pairs, associated with the photonic, atomic and phononic branches. The phononic mode eigenvalues remain constant and imaginary for all the values of atom-photon coupling strength. Moreover, for the parameter regime we have considered, the fluorescence spectra and the homodyne spectra exhibit normal mode splitting which shows the coherent energy exchange between the different modes (photonic mode, Bogoliubov mode and the mechanical mode) of the system. Both the spectra display striking behaviour in the vicinity of critical point. The most interesting observation in the fluorescence spectrum is the asymmetric coherent energy exchange between the three modes of the system in the superradiant phase. Such an asymmetry arises in the presence of optomechanical interaction as a result of Bloch-siegert shift and this asymmetry increases as the atom-photon coupling strength increases. Moreover, the homodyne spectra of the output cavity field can be used to monitor the atom-cavity field entanglement.

\section{Acknowledgements}

Neha Aggarwal and A. Bhattacherjee acknowledge financial support from the Department of Science and Technology, New Delhi for financial assistance vide grant SR/S2/LOP-0034/2010. Sonam Mahajan acknowledges University of Delhi for the University Teaching Assistantship.

\section{Appendix A}

The coefficients in eqn.(\ref{ham4}) are given as follows:

\begin{equation}
x_{1}=\omega_{0}-\frac{2\sqrt{2}\alpha_{ss}Re[\gamma_{ss}]g}{N\sqrt{x_{8}(1+\mu)}},
\end{equation}

\begin{equation}
x_{2}=g\sqrt{\frac{x_{8}}{N}}\left(1-\frac{2\alpha_{ss}^{2}}{x_{8}N(1+\mu)} \right),
\end{equation}

\begin{equation}
x_{3}=\omega_{c}\eta_{0}Re[\gamma_{ss}],
\end{equation}

\begin{equation}
x_{4}=-2g\alpha_{ss}Re[\gamma_{ss}]\sqrt{\frac{x_{8}}{N}}\left[\frac{1}{x_{8}\sqrt{2N(1+\mu)}}-\frac{\alpha_{ss}^{2}}{4x_{8}^{2}\left\lbrace \frac{N(1+\mu)}{2} \right\rbrace^{3/2} } \right],
\end{equation}

\begin{equation}
x_{5}=\frac{\omega_{0}\alpha_{ss}}{\sqrt{\frac{N(1+\mu)}{2}}}+2gRe[\gamma_{ss}]\sqrt{\frac{x_{8}}{N}}-\frac{4Re[\gamma_{ss}]\alpha_{ss}^{2}g}{N(1+\mu)\sqrt{x_{8}N}},
\end{equation}

\begin{equation}
x_{6}=\omega_{m}Re[\beta_{ss}]+\omega_{c}\eta_{0}(Re[\gamma_{ss}])^{2},
\end{equation}

\begin{equation}
x_{7}=\omega_{1} Re[\gamma_{ss}]+\frac{2g\alpha_{ss}\sqrt{2 x_{8}}}{N\sqrt{1+\mu}},
\end{equation}

where,

\begin{equation}
x_{8}=N-\frac{2\alpha_{ss}^{2}}{N(1+\mu)}.
\end{equation}

\section{Appendix B}

The cavity input noise operators satisfy the following correlations \citep{mancini,vitali,giovannetti,vitali1} :

\begin{equation}
\langle c_{in}(t),c_{in}(t')\rangle =\langle c_{in}^{\dagger}(t),c_{in}(t')\rangle =0,
\end{equation}

\begin{equation}
\langle c_{in}(t),c_{in}^{\dagger}(t')\rangle =\delta(t-t').
\end{equation}

The noise operator due to the Brownian motion of the mirror follows the following correlation \citep{giovannetti,vitali1}:

\begin{equation}
<W(t)W(t')>=\frac{\gamma_{m}}{\omega_{m}}\int \frac{d \omega}{2\pi}e^{-i\omega(t-t')}\omega \left[ 1+\coth\left(\frac{\hbar \omega}{2k_{B}T} \right)\right],
\end{equation}

where $k_{B}$ is the Boltzmann constant and $T$ represents the finite temperature of the bath connected to the cantilever. Brownian noise is the random thermal noise which arises due to the stochastic motion of the mechanical mirror and is non-Markovian in nature.

The frequency space equivalents of the input correlations are \citep{giovannetti}:

$\langle c_{in}(\omega),c_{in}^{\dagger}(\omega') \rangle=2 \pi \delta(\omega-\omega')$,
$\langle c_{in}(\omega),c_{in}(\omega')\rangle =0$ and $\langle c_{in}^{\dagger}(\omega),c_{in}(\omega')\rangle =0$.

Also the correlation function for the Brownian noise operator in Fourier space is given as \citep{giovannetti}:

$\langle W(\omega) W(\omega') \rangle=2 \pi \frac{\gamma_{m}}{\omega_{m}} \omega \left[1+\coth \left ( {\frac{\hbar \omega}{2 k_{B} T}} \right )\right] \delta(\omega+\omega')$.

\section{Appendix C}

The expressions for the coefficients used in eqn.(\ref{eq12}) are:

\begin{equation}
A_{1}(\omega)=(x_{1}^{2}-\omega^{2}+4x_{1}x_{4})[{\gamma_{c}+i(\omega_{1} -\omega)}(\omega^{2}-\omega_{m}^{2}+i\omega \gamma_{m})+2ix_{3}^{2}\omega_{m}]-2ix_{1}x_{2}^{2}(\omega^{2}-\omega_{m}^{2}+i\omega \gamma_{m}),
\end{equation}

\begin{equation}
A_{2}(\omega)=(x_{1}^{2}-\omega^{2}+4x_{1}x_{4})[{\gamma_{c}-i(\omega_{1} +\omega)}(\omega^{2}-\omega_{m}^{2}+i\omega \gamma_{m})-2ix_{3}^{2}\omega_{m}]+2ix_{1}x_{2}^{2}(\omega^{2}-\omega_{m}^{2}+i\omega \gamma_{m}),
\end{equation}

\begin{equation}
A_{3}(\omega)=x_{1}x_{2}^{2}(\omega^{2}-\omega_{m}^{2}+i\omega \gamma_{m})-x_{3}^{2}\omega_{m}(x_{1}^{2}-\omega^{2}+4x_{1}x_{4}),
\end{equation}

\begin{equation}
A_{4}(\omega)=x_{3}\omega_{m}(x_{1}^{2}-\omega^{2}+4x_{1}x_{4}),
\end{equation}

\begin{equation}
A_{5}(\omega)=(x_{1}^{2}-\omega^{2}+4x_{1}x_{4})(\omega^{2}-\omega_{m}^{2}+i\omega \gamma_{m}).
\end{equation}

The other coefficients used in eqns.(\ref{eq13}) and (\ref{eq14}) are given as follows:

\begin{equation}
B_{1}(\omega)=\left[ \left\lbrace \gamma_{c}+i(\omega -\omega_{c})\right\rbrace \left\lbrace \gamma_{c}+i(\omega +\omega_{c}) \right\rbrace(\omega^{2}-\omega_{0}^{2})+4\omega_{0}^{2}g^{2}\right],
\end{equation}

\begin{equation}
B_{2}(\omega)=\frac{8\pi\gamma_{m}\omega}{\omega_{m}}\left\lbrace 1+\coth\left(\frac{\hbar \omega}{2k_{B}T}\right)\right\rbrace \left[A_{4}(-\omega)A_{3}(-\omega)A_{4}^{\dagger}(-\omega)A_{3}^{\dagger}(-\omega) \right],
\end{equation}

\begin{equation}
B_{3}(\omega)=\frac{4i \pi\gamma_{m}\omega}{\omega_{m}}\left\lbrace 1+\coth\left(\frac{\hbar \omega}{2k_{B}T}\right)\right\rbrace \left[A_{4}(-\omega)A_{2}(-\omega)A_{4}^{\dagger}(-\omega)A_{3}^{\dagger}(-\omega) \right],
\end{equation}

\begin{equation}
B_{4}(\omega)=\frac{-4i \pi\gamma_{m}\omega}{\omega_{m}}\left\lbrace 1+\coth\left(\frac{\hbar \omega}{2k_{B}T}\right)\right\rbrace \left[A_{4}(-\omega)A_{3}(-\omega)A_{4}^{\dagger}(-\omega)A_{2}^{\dagger}(-\omega) \right],
\end{equation}

\begin{equation}
B_{5}(\omega)=\frac{2\pi\gamma_{m}\omega}{\omega_{m}}\left\lbrace 1+\coth\left(\frac{\hbar \omega}{2k_{B}T}\right)\right\rbrace \left[A_{4}(-\omega)A_{2}(-\omega)A_{4}^{\dagger}(-\omega)A_{2}^{\dagger}(-\omega) \right],
\end{equation}

\begin{equation}
B_{6}(\omega)=16\pi \gamma_{c}\left[A_{5}(-\omega)A_{3}(-\omega)A_{5}^{\dagger}(-\omega)A_{3}^{\dagger}(-\omega)\right].
\end{equation}

\section{Appendix D}

The values of the coefficients used in eqn.(\ref{eq15}) are:

\begin{equation}
C_{1}(\omega)=-\frac{16i\pi\omega_{0}g^{2}\gamma_{c}^{2}C_{5}(\omega)}{C_{6}(\omega)C_{6}(-\omega)}+\frac{8i\pi\omega_{0}g^{2}\gamma_{c}}{C_{6}(-\omega)},
\end{equation}

\begin{equation}
C_{2}(\omega)=2\pi +8\pi \frac{\gamma_{c}^{2}C_{5}(\omega)C_{5}^{\dagger}(\omega)}{C_{6}(\omega)C_{6}^{\dagger}(\omega)}-4\pi \gamma_{c}\frac{C_{5}^{\dagger}(\omega)}{C_{6}^{\dagger}(\omega)}-4\pi \gamma_{c}\frac{C_{5}(\omega)}{C_{6}(\omega)},
\end{equation}

\begin{equation}
C_{3}(\omega)=\frac{32\pi \omega_{0}^{2}g^{4}\gamma_{c}^{2}}{C_{6}^{\dagger}(-\omega)C_{6}(-\omega)},
\end{equation}

\begin{equation}
C_{4}(\omega)=\frac{16i\pi \omega_{0}g^{2}\gamma_{c}^{2}C_{5}^{\dagger}(\omega)}{C_{7}(\omega)C_{7}(-\omega)}-\frac{8i\pi \omega_{0}g^{2}\gamma_{c}}{C_{7}(\omega)},
\end{equation}

\begin{equation}
C_{5}(\omega)=\left[\left\lbrace \gamma_{c}-i(\omega +\omega_{c})\right\rbrace(\omega^{2}-\omega_{0}^{2})-2i\omega_{0}g^{2}\right],
\end{equation}

\begin{equation}
C_{6}(\omega)=\left[\left\lbrace \gamma_{c}-i(\omega -\omega_{c})\right\rbrace \left\lbrace \gamma_{c}-i(\omega +\omega_{c})\right\rbrace(\omega^{2}-\omega_{0}^{2})\right]+4\omega_{0}^{2}g^{2},
\end{equation}

\begin{equation}
C_{7}(\omega)=\left[\left\lbrace \gamma_{c}-i(\omega +\omega_{c})\right\rbrace \left\lbrace \gamma_{c}+i(-\omega +\omega_{c})\right\rbrace(\omega^{2}-\omega_{0}^{2})\right]+4\omega_{0}^{2}g^{2}.
\end{equation}

The other values of the coefficients used in eqn.(\ref{eq16}) are given as:

\begin{equation}
D_{1}(\omega)=\frac{2\gamma_{c}}{\left[A_{1}(-\omega)A_{2}(-\omega)-4(A_{3}(-\omega))^{2}\right]}\left\lbrace \frac{\left[\frac{2\pi\gamma_{m}}{\omega_{m}}\omega \left[ 1+\coth\left\lbrace \frac{\hbar \omega}{2k_{B}T}\right\rbrace\right] D_{5}(\omega)+D_{6}(\omega)\right]}{\left[A_{1}(\omega)A_{2}(\omega)-4(A_{3}(\omega))^{2}\right]}-D_{7}(\omega)\right\rbrace ,
\end{equation}

\begin{equation}
D_{2}(\omega)=\frac{2\gamma_{c}\left\lbrace \frac{2\pi\gamma_{m}}{\omega_{m}}\omega \left[ 1+\coth\left\lbrace \frac{\hbar \omega}{2k_{B}T}\right\rbrace\right] D_{8}(\omega)+D_{9}(\omega)\right\rbrace }{\left[A_{1}^{\dagger}(-\omega)A_{2}^{\dagger}(-\omega)-4(A_{3}^{\dagger}(-\omega))^{2}\right]\left[A_{1}(-\omega)A_{2}(-\omega)-4(A_{3}(-\omega))^{2}\right]},
\end{equation}

\begin{equation}
D_{3}(\omega)=2\pi +\frac{2\gamma_{c} \left\lbrace \frac{2\pi\gamma_{m}}{\omega_{m}}\omega \left[ 1+\coth\left\lbrace \frac{\hbar \omega}{2k_{B}T}\right\rbrace\right]D_{10}(\omega)+D_{11}(\omega)\right\rbrace }{\left[A_{1}^{\dagger}(\omega)A_{2}^{\dagger}(\omega)-4(A_{3}^{\dagger}(\omega))^{2}\right]\left[A_{1}(\omega)A_{2}(\omega)-4(A_{3}(\omega))^{2}\right]}-D_{12}(\omega)-D_{13}(\omega),
\end{equation}

\begin{equation}
D_{4}(\omega)=\frac{2\gamma_{c}}{\left[A_{1}^{\dagger}(-\omega)A_{2}^{\dagger}(-\omega)-4(A_{3}^{\dagger}(-\omega))^{2}\right]}\left\lbrace \frac{\left[\frac{2\pi\gamma_{m}}{\omega_{m}}\omega \left[ 1+\coth\left\lbrace \frac{\hbar \omega}{2k_{B}T}\right\rbrace\right] D_{14}(\omega)-D_{15}(\omega)\right]}{\left[A_{1}^{\dagger}(\omega)A_{2}^{\dagger}(\omega)-4(A_{3}^{\dagger}(\omega))^{2}\right]}+D_{16}(\omega)\right\rbrace ,
\end{equation}

\begin{equation}
D_{5}(\omega)=\left[2A_{4}(\omega)A_{3}(\omega)+iA_{4}(\omega)A_{2}(\omega)\right] \left[ 2A_{4}(-\omega)A_{3}(-\omega)+iA_{4}(-\omega)A_{2}(-\omega)\right],
\end{equation}

\begin{equation}
D_{6}(\omega)=8i\pi \gamma_{c}A_{5}(\omega)A_{2}(\omega)A_{5}(-\omega)A_{3}(-\omega),
\end{equation}

\begin{equation}
D_{7}(\omega)=4i\pi A_{5}(-\omega)A_{3}(-\omega),
\end{equation}

\begin{equation}
D_{8}(\omega)=\left[2A_{4}^{\dagger}(-\omega)A_{3}^{\dagger}(-\omega)-iA_{4}^{\dagger}(-\omega)A_{2}^{\dagger}(-\omega)\right]\left[ 2A_{4}(-\omega)A_{3}(-\omega)+iA_{4}(-\omega)A_{2}(-\omega)\right],
\end{equation}

\begin{equation}
D_{9}(\omega)=16\pi \gamma_{c}A_{5}^{\dagger}(-\omega)A_{3}^{\dagger}(-\omega)A_{5}(-\omega)A_{3}(-\omega),
\end{equation}

\begin{equation}
D_{10}(\omega)=\left[2A_{4}^{\dagger}(\omega)A_{3}^{\dagger}(\omega)-iA_{4}^{\dagger}(\omega)A_{2}^{\dagger}(\omega)\right]\left[ 2A_{4}(\omega)A_{3}(\omega)+iA_{4}(\omega)A_{2}(\omega)\right],
\end{equation}

\begin{equation}
D_{11}(\omega)=4 \pi \gamma_{c}A_{5}^{\dagger}(\omega)A_{2}^{\dagger}(\omega)A_{5}(\omega)A_{2}(\omega),
\end{equation}

\begin{equation}
D_{12}(\omega)=\frac{4 \pi \gamma_{c}A_{5}(\omega)A_{2}(\omega)}{\left[A_{1}(\omega)A_{2}(\omega)-4(A_{3}(\omega))^{2} \right]},
\end{equation}

\begin{equation}
D_{13}(\omega)=\frac{4 \pi \gamma_{c}A_{5}^{\dagger}(\omega)A_{2}^{\dagger}(\omega)}{\left[A_{1}^{\dagger}(\omega)A_{2}^{\dagger}(\omega)-4(A_{3}^{\dagger}(\omega))^{2} \right]},
\end{equation}

\begin{equation}
D_{14}(\omega)=\left[2A_{4}^{\dagger}(-\omega)A_{3}^{\dagger}(-\omega)-iA_{4}^{\dagger}(-\omega)A_{2}^{\dagger}(-\omega)\right]\left[ 2A_{4}^{\dagger}(\omega)A_{3}^{\dagger}(\omega)-iA_{4}^{\dagger}(\omega)A_{2}^{\dagger}(\omega)\right],
\end{equation}

\begin{equation}
D_{15}(\omega)=8i \pi \gamma_{c}A_{5}^{\dagger}(-\omega)A_{3}^{\dagger}(-\omega)A_{5}^{\dagger}(\omega)A_{2}^{\dagger}(\omega),
\end{equation}

\begin{equation}
D_{16}(\omega)=4i \pi A_{5}^{\dagger}(-\omega)A_{3}^{\dagger}(-\omega).
\end{equation}

\end{document}